\documentclass[sigconf,authorversion,nonacm]{acmart}

\usepackage{csquotes}
\usepackage{subcaption}
\usepackage{hyperref}
\usepackage{lscape}
\AtBeginDocument{%
  }

\begin{document}

%%
%% The "title" command has an optional parameter,
%% allowing the author to define a "short title" to be used in page headers.
\title{The Monetisation of Toxicity: Analysing YouTube Content Creators and Controversy-Driven Engagement}

%%
%% The "author" command and its associated commands are used to define
%% the authors and their affiliations.
%% Of note is the shared affiliation of the first two authors, and the
%% "authornote" and "authornotemark" commands
%% used to denote shared contribution to the research.
\author{Thales Bertaglia}
\email{t.f.costabertaglia@uu.ml}

\author{Catalina Goanta}
\email{e.c.goanta@uu.nl}
\affiliation{%
  \institution{Utrecht University}
  \city{Utrecht}
  \country{the Netherlands}
}

\author{Adriana Iamnitchi}
\affiliation{%
  \institution{Maastricht University}
  \city{Maastricht}
  \country{the Netherlands}}
\email{a.iamnitchi@maastrichtuniversity.nl}

%%
%% By default, the full list of authors will be used in the page
%% headers. Often, this list is too long, and will overlap
%% other information printed in the page headers. This command allows
%% the author to define a more concise list
%% of authors' names for this purpose.
\renewcommand{\shortauthors}{Bertaglia et al.}

%%
%% The abstract is a short summary of the work to be presented in the
%% article.
\begin{abstract}
YouTube is a major social media platform that plays a significant role in digital culture, with content creators at its core. These creators often engage in controversial behaviour to drive engagement, which can foster toxicity. This paper presents a quantitative analysis of controversial content on YouTube, focusing on the relationship between controversy, toxicity, and monetisation. We introduce a curated dataset comprising 20 controversial YouTube channels extracted from Reddit discussions, including 16,349 videos and more than 105 million comments. We identify and categorise monetisation cues from video descriptions into various models, including affiliate marketing and direct selling, using lists of URLs and keywords. Additionally, we train a machine learning model to measure the toxicity of comments in these videos. Our findings reveal that while toxic comments correlate with higher engagement, they negatively impact monetisation, indicating that controversy-driven interaction does not necessarily lead to financial gain. We also observed significant variation in monetisation strategies, with some creators showing extensive monetisation despite high toxicity levels. Our study introduces a curated dataset, lists of URLs and keywords to categorise monetisation, a machine learning model to measure toxicity, and is a significant step towards understanding the complex relationship between controversy, engagement, and monetisation on YouTube. The lists used for detecting and categorising monetisation cues are available on \url{https://github.com/thalesbertaglia/toxmon}. 
\end{abstract}

% \begin{CCSXML}
% <ccs2012>
%    <concept>
%        <concept_id>10002951.10003260.10003272.10003276</concept_id>
%        <concept_desc>Information systems~Social advertising</concept_desc>
%        <concept_significance>500</concept_significance>
%        </concept>
%  </ccs2012>
% \end{CCSXML}

% \ccsdesc[500]{Information systems~Social advertising}

% %%
% %% Keywords. The author(s) should pick words that accurately describe
% %% the work being presented. Separate the keywords with commas.
% \keywords{content monetization, content creators, influencers, influencer marketing, youtube, toxicity}

\maketitle

\section{Introduction and Related Work}
YouTube is one of the most popular social media platforms, yet it remains surprisingly understudied within computational academic research. Content creators are central to the YouTube ecosystem and have a significant influence, especially on vulnerable individuals, such as children. These creators often find themselves involved in controversy and are often the target of hate, leading to negative repercussions that can propagate across social media.

Research indicates that controversial behaviour can be strategically used to generate engagement~\cite{bertaglia2021clout}. Therefore, despite its adverse consequences, controversy and toxicity can serve as viable monetisation strategies. However, there is a gap in research exploring the relationship between monetisation, controversy, and toxicity on YouTube. This paper aims to fill this gap by investigating how these elements interact and affect the overall dynamics of content creation on the platform.

Existing studies have explored various aspects of YouTube, such as toxicity and hate speech~\cite{bertaglia2021abusive,gupta2023hateful,Papadamou_2021,alshamrani2020investigating}, sexism~\cite{bertaglia2023sexism,amarasekara2019exploring,szostak2013girls}, and the practices of specific content creators~\cite{wotanis2014performing,dawson2021trisha,geusens2023cancelled}. Additionally, research has mapped out monetisation models at scale, highlighting the various strategies employed by creators to generate revenue~\cite{hua2022characterizing,rieder2023making,mathurEndorsementsSocialMedia2018}.

Studies have mapped out the monetisation landscape, revealing a broad range of strategies beyond platform-provided tools, including external monetisation methods that have become increasingly prevalent~\cite{hua2022characterizing}. \citet{hua2022characterizing} show that problematic channels producing content related to Alt-lite, Alt-right, and the Manosphere employ a diverse set of alternative monetisation strategies more frequently than other channels, complicating YouTube's role as a gatekeeper. Moreover, the economic pressures exerted by YouTube's platformed media system have driven creators to adopt sophisticated monetisation and networking strategies, such as linking to external platforms in video descriptions to expand their income streams~\cite{rieder2023making}. 

Beyond traditional social media platforms such as YouTube, recent studies have also investigated Blockchain-based Online Social Media (BOSM) platforms, which provide a decentralised alternative to conventional social networks. BOSMs such as Steemit and Yup incorporate blockchain technology to offer transparent and auditable reward systems for user-generated content~\cite{guidi2020graph,guidi2022reward}. These platforms present unique dynamics, where users are incentivised to be socially active through economic rewards, yet the distribution of wealth and social engagement can vary significantly, often favouring the most active content producers~\cite{guidi2020rewarding}. 

Building on these insights, we present a quantitative analysis of controversial YouTube content, focusing on the evolution of monetisation strategies, variations in engagement rates, and the prevalence of toxic comments. By analysing data from video descriptions, we identify and categorise monetisation strategies and examine how controversies influence engagement. We further explore the relationship between increasing comment toxicity and its impact on monetisation, providing a comprehensive understanding of how controversy-driven user interaction contributes to the rise of harmful communication.

\section{Dataset}
Our experiments rely on a curated dataset of controversial YouTubers sourced from Reddit discussions. We constructed the sample by selecting content creators mentioned in subreddits discussing YouTube controversies~\cite{reddit}. We relied on Reddit threads from two subreddits focused on discussing \textit{internet drama}, r/YouTubeDrama\footnote{\url{https://www.reddit.com/r/youtubedrama}} and r/InternetDrama\footnote{\url{https://www.reddit.com/r/internetdrama}}. Internet drama refers to contentious or sensational discourse, events, or personalities within online communities, often characterised by disputes, controversies, or high emotional engagement~\cite{lewis2022platform}. r/YouTubeDrama focuses on controversies and discussions surrounding YouTube personalities, channels, and content, providing a space for community-driven discourse. r/InternetDrama, on the other hand, broadly includes any form of drama across the entire internet. We collected data (thread text and comments) from the top 1000 threads of all time, sorted by \textit{Hot} from each subreddit; \textit{Hot} is a Reddit sorting metric that prioritises threads based on a combination of their recency and the rate at which they are receiving upvotes, comments, and overall engagement, resulting in a mix of recent and older threads -- spanning a comprehensive range of potentially controversial content. We collected all data using the official Reddit API. We only kept threads with at least five comments because we noticed that threads with low engagement (usually at most one comment)  tended to be self-promotion (YouTubers posting their own content). We collected all Reddit data in November 2023. 

After collecting the threads and comments, we processed them to identify the YouTube channels users were discussing. Some threads directly embedded YouTube content, meaning they included specific YouTube videos within the post itself through a feature that allows users to watch videos directly on Reddit. We also extracted all YouTube URLs from the text of threads and comments. When URLs contained videos, we used the YouTube API to retrieve channel information. Many Reddit comments talk about YouTubers using their names or channel names, particularly for very popular ones. To identify such mentions, we employed Named Entity Recognition (NER) using the \textit{en\_core\_web\_trf}\footnote{\url{https://spacy.io/models/en}} model from spaCy~\cite{honnibal2020spacy}. Finally, we aggregated all mentions of YouTube channels and manually selected a subset to include in our dataset. This curation step ensured our dataset focused on relevant discussions and channels, providing a solid foundation for our analysis.

We classified controversial YouTube channels into two distinct conceptual categories based on their nature and patterns of controversy:

\begin{enumerate}
    \item \textbf{Consistent Controversy Channels (Cons.)}: These channels are often involved in controversies. They are well-known, popular channels repeatedly involved in scandals. Examples include high-profile creators like James Charles and Logan Paul, who have been involved in various controversies over time. Comments mostly referred to these channels by name, so we identified them primarily through NER.
    \item \textbf{Spike Controversy Channels}: This category contains channels discussed on recent \textit{Hot} threads (by the time of our data collection) due to their involvement in a particular controversy or because they are newer, smaller channels that have exhibited controversial behaviour. One example of a channel in this category is \textit{The Completionist}, who was accused of withholding over \$600,000 in charity donations in November 2023~\cite{dexerto2023}. 
\end{enumerate}

To compile a representative sample, we selected 20 channels in total, evenly split between the two categories. We sorted all mentioned channels by the frequency of mentions, then reviewed this list to choose the ten channels that most accurately represented each category. This methodological approach allowed us to capture a diverse set of controversial channels, encompassing both those with a consistent history of controversies and those whose controversial status may be temporary or tied to specific recent events.

With our curated list of channels, we used the official YouTube API to collect all relevant data and metadata about the channels, their videos, and comments. We did not collect comment replies and focused exclusively on collecting top-level comments, as they are more likely to engage directly with the video's content or the YouTuber themselves. We collected the data for channels and videos in November 2023. However, due to the vast volume of comments and the limitations imposed by the API, the collection of comment data was extended over several phases, spanning from November 2023 to January 2024. \autoref{tab:channels_desc} summarises the general statistics of the channels included in the dataset. In total, our dataset contains 16,349 videos and 105,854,713 comments.

The channel names retrieved through the API at the time of collection may not match those currently displayed on the YouTube interface, as creators can change them. For example, the channel identified as \textit{blndsundoll4mj} corresponds to Trisha Paytas's channel, a creator frequently involved in various controversies and whose content has sparked debates on mental health, body image, and the boundaries of content creation~\cite{dawson2021trisha,berraf2022instagram}.

\begin{table}
\tiny
\centering
\caption{Descriptive statistics of the channels in the dataset.}
\label{tab:channels_desc}
\begin{tabular}{lcccccc}
\toprule
\bfseries Channel & \bfseries Category & \bfseries Views (M) & \bfseries Subs (M) & \bfseries Videos & \bfseries Age (years) \\
\midrule
SSSniperWolf & Spike & 24,182.33 & 34.10 & 3,452 & 11.03 \\
James Charles & Cons. & 4,213.73 & 23.90 & 538 & 8.16 \\
Logan Paul & Cons. & 5,997.71 & 23.60 & 716 & 8.42 \\
Jake Paul & Cons. & 7,438.79 & 20.50 & 1,148 & 10.37 \\
JennaMarbles & Cons. & 1,816.66 & 19.70 & 250 & 13.95 \\
shane & Cons. & 4,346.19 & 19.10 & 571 & 18.36 \\
David Dobrik & Cons. & 7,208.18 & 17.70 & 536 & 9.11 \\
jeffreestar & Cons. & 2,585.32 & 15.80 & 433 & 17.96 \\
Colleen Ballinger & Spike & 1,918.35 & 8.41 & 1,091 & 17.50 \\
The Gabbie Show & Cons. & 134.20 & 5.15 & 46 & 9.97 \\
blndsundoll4mj & Cons. & 953.68 & 5.10 & 2,259 & 17.07 \\
boogie2988 & Spike & 928.07 & 4.02 & 2,339 & 17.82 \\
Nikocado Avocado & Cons. & 778.69 & 3.73 & 706 & 9.68 \\
The Completionist & Spike & 335.85 & 1.62 & 699 & 12.07 \\
iilluminaughtii & Spike & 254.15 & 1.31 & 633 & 10.73 \\
JessiSmiles & Spike & 107.47 & 0.99 & 189 & 10.52 \\
Yumi King & Spike & 147.38 & 0.85 & 1,087 & 10.77 \\
nickisnotgreen & Spike & 48.42 & 0.69 & 128 & 7.33 \\
Life Plus Cindy & Spike & 3.76 & 0.02 & 226 & 2.82 \\
lil lunchbox & Spike & 1.07 & 0.01 & 308 & 6.92 \\
\bottomrule
\end{tabular}
\end{table}

\begin{figure*}[htbp]
    \centering
    \begin{subfigure}[b]{0.45\linewidth}
        \includegraphics[width=\linewidth]{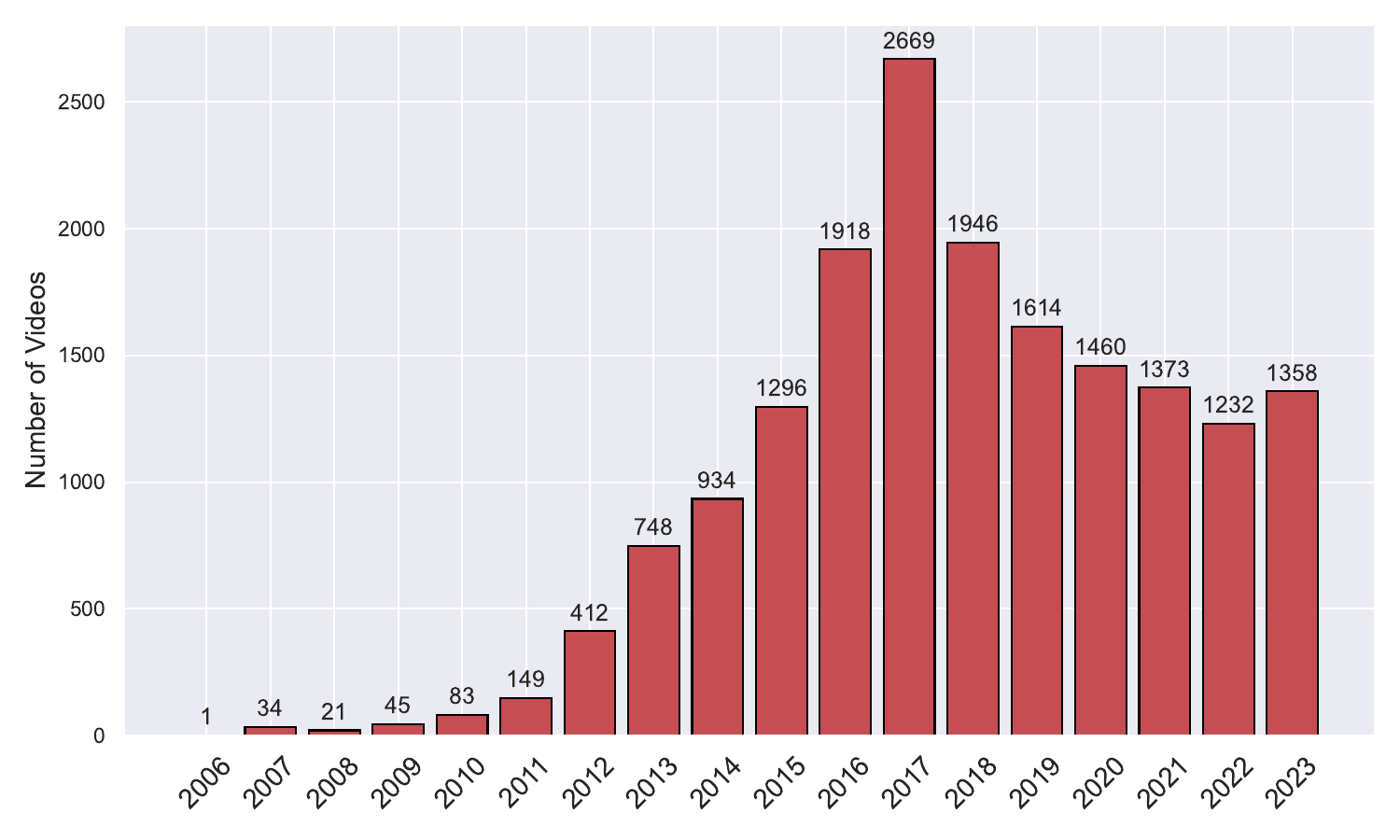}
        \caption{Number of videos published per year considering all channels in the dataset.}
        \label{fig:sub_videos_per_year}
    \end{subfigure}
    \hspace{0.05\linewidth}
    \begin{subfigure}[b]{0.45\linewidth}
        \includegraphics[width=\linewidth]{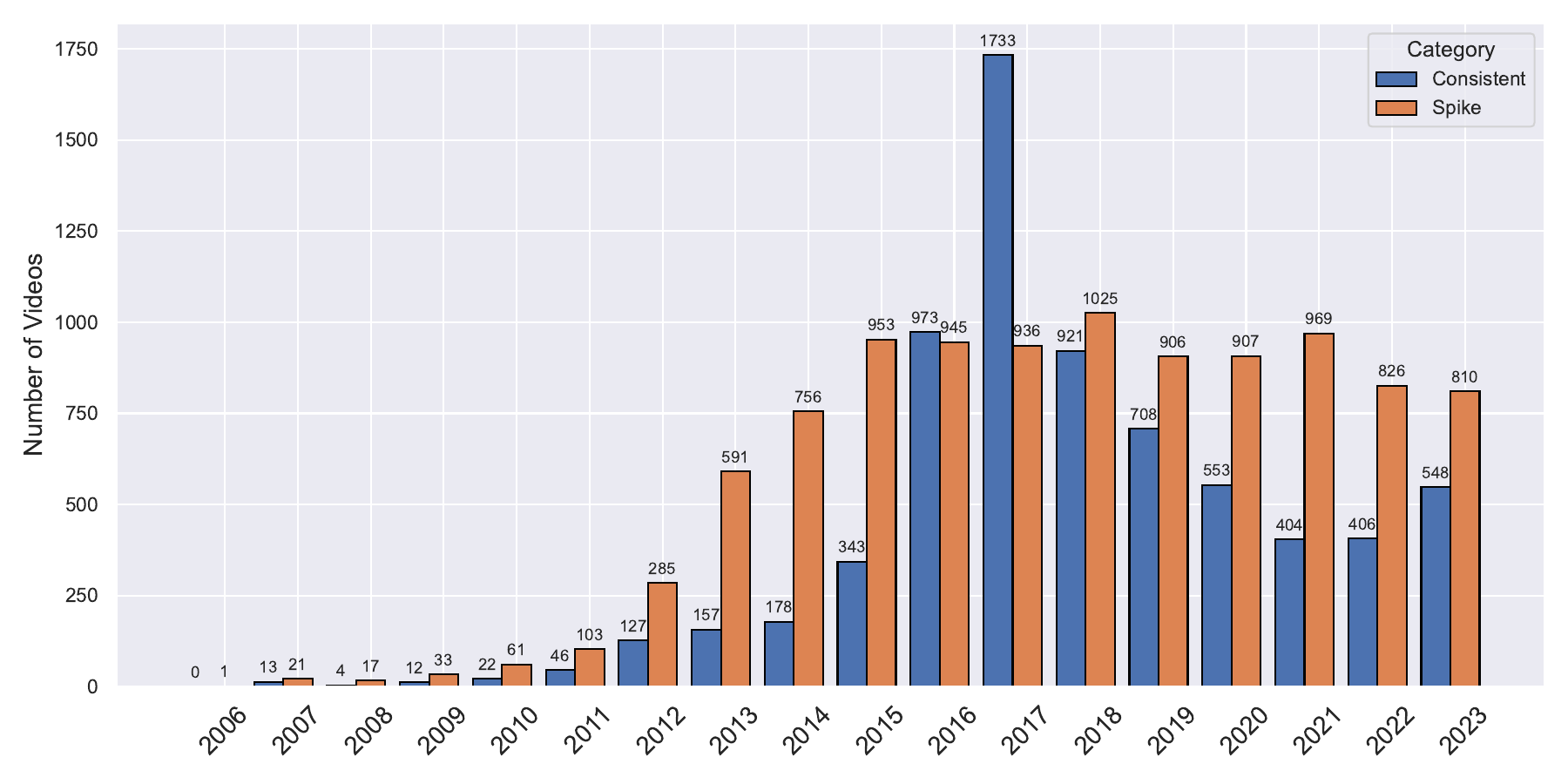}
        \caption{Number of videos published per year split by controversy category.}
        \label{fig:sub_videos_per_year_per_category}
    \end{subfigure}
    \caption{Videos published per year in the dataset.}
    \label{fig:videos_per_year_combined}
\end{figure*}

The distribution of subscriber counts and views across our dataset illustrates a wide range of channel sizes, including highly popular creators with more than 30 million subscribers to smaller channels with approximately 100,000 subscribers. The \textit{Consistent} category predominantly comprises larger channels, aligning with how we define it. Except for \textit{Life Plus Cindy}, all channels have been active for over six years, indicating they are established content creators. There is a significant variation in the number of videos each channel has published; ~\autoref{fig:videos_per_year_combined} presents the number of videos published per year across the entire dataset and split by controversy category.

% \begin{figure*}[htbp]
%     \centering
%     \includegraphics[width=\linewidth]{figures/videos_per_year}
%     \caption{Number of videos published per year considering all channels in the dataset.}
%     \label{fig:videos_per_year}
% \end{figure*}

% \begin{figure*}[htbp]
% \centering
% \includegraphics[width=\linewidth]{figures/videos_per_year_per_category}
% \caption{Number of videos published per year split by drama category.}
% \label{fig:videos_per_year_per_category}
% \end{figure*}

Finally, ~\autoref{tab:videos_stats} provides a statistical summary of aggregated video metrics, including views, likes, and comment counts, for each channel, presented in terms of mean and standard deviation values. It is important to note that the number of comments reported by the API includes both comments and their replies, whereas we only collected top-level comments. In practice, the comment counts from the API are higher than the actual number of comments collected, since we did not collect replies.

\begin{table*}
% \begin{table}
\small
\centering
\caption{Statistical summary of aggregated video metrics (view, like, and comment counts) for each channel presented as mean and standard deviation values.}
\label{tab:videos_stats}
\begin{tabular}{lcccc}
\toprule
\bfseries Channel & \bfseries Views & \bfseries Likes & \bfseries Comments \\
\midrule
SSSniperWolf & 7,045.06k $\pm$ 8,112.70k & 147.05k $\pm$ 162.36k & 10.18k $\pm$ 14.70k \\
James Charles & 7,764.84k $\pm$ 8,547.91k & 327.99k $\pm$ 366.15k & 23.52k $\pm$ 48.72k \\
Logan Paul & 8,405.08k $\pm$ 13,477.09k & 310.24k $\pm$ 287.94k & 30.06k $\pm$ 72.57k \\
Jake Paul & 6,518.39k $\pm$ 12,409.05k & 151.57k $\pm$ 183.78k & 20.39k $\pm$ 43.64k \\
JennaMarbles & 7,249.94k $\pm$ 5,517.20k & 267.22k $\pm$ 119.54k & 18.88k $\pm$ 12.68k \\
shane & 7,573.17k $\pm$ 7,121.58k & 282.71k $\pm$ 285.67k & 31.35k $\pm$ 39.22k \\
David Dobrik & 13,457.43k $\pm$ 6,709.35k & 407.11k $\pm$ 248.70k & 8.74k $\pm$ 12.21k \\
jeffreestar & 5,928.46k $\pm$ 5,956.43k & 197.84k $\pm$ 213.76k & 28.15k $\pm$ 63.34k \\
Colleen Ballinger & 1,766.73k $\pm$ 2,903.07k & 54.68k $\pm$ 82.51k & 5.08k $\pm$ 10.79k \\
The Gabbie Show & 2,936.00k $\pm$ 4,295.07k & 121.66k $\pm$ 170.94k & 18.99k $\pm$ 28.27k \\
blndsundoll4mj & 434.96k $\pm$ 874.98k & 9.26k $\pm$ 16.00k & 1.76k $\pm$ 2.94k \\
boogie2988 & 396.99k $\pm$ 1,323.02k & 12.43k $\pm$ 23.26k & 2.15k $\pm$ 4.17k \\
Nikocado Avocado & 1,102.42k $\pm$ 1,825.45k & 14.95k $\pm$ 21.43k & 4.65k $\pm$ 8.46k \\
The Completionist & 480.45k $\pm$ 690.61k & 13.47k $\pm$ 11.72k & 1.54k $\pm$ 1.39k \\
iilluminaughtii & 401.53k $\pm$ 441.08k & 16.60k $\pm$ 17.74k & 2.05k $\pm$ 2.72k \\
JessiSmiles & 566.23k $\pm$ 734.84k & 24.55k $\pm$ 33.69k & 1.99k $\pm$ 3.16k \\
Yumi King & 141.51k $\pm$ 462.06k & 2.94k $\pm$ 5.15k & 0.17k $\pm$ 0.23k \\
nickisnotgreen & 380.25k $\pm$ 450.64k & 24.23k $\pm$ 25.11k & 1.53k $\pm$ 2.26k \\
Life Plus Cindy & 16.66k $\pm$ 14.52k & 1.30k $\pm$ 0.93k & 0.31k $\pm$ 0.39k \\
lil lunchbox & 3.48k $\pm$ 5.98k & 0.05k $\pm$ 0.05k & 0.05k $\pm$ 0.05k \\
\midrule
\bfseries Consistent & 4,967.97k $\pm$ 8,719.13k & 159.84k $\pm$ 241.59k & 14.69k $\pm$ 39.17k \\
\bfseries Spike & 2,848.63k $\pm$ 5,849.75k & 63.66k $\pm$ 118.90k & 4.96k $\pm$ 10.48k \\
\midrule
\bfseries All & 3,740.71k $\pm$ 7,273.73k & 104.15k $\pm$ 187.10k & 9.05k $\pm$ 27.06k \\
\bottomrule
\end{tabular}
% \end{table}
\end{table*}

\textit{Consistent} channels display significantly higher engagement metrics, with an average number of comments approximately three times greater than \textit{Spike}. This discrepancy aligns with our expectations, given that \textit{Consistent} channels are generally larger, except for \textit{SSSniperWolf}, which has the most subscribers of all channels in the dataset. Conversely, Spike Controversy channels show greater activity levels, as illustrated in ~\autoref{fig:videos_per_year_combined}, indicating that channels within this category publish more videos yearly.

\section{Detecting Monetisation Cues}
Identifying the monetisation models YouTubers employ is essential for understanding how controversial content creators generate revenue. Since the YouTube API does not directly provide details about sponsorships or monetisation strategies in video metadata, we analysed video descriptions to detect monetisation cues. Creators often mention their sponsors, affiliate links, or other revenue sources in these descriptions, such as promoting a specific brand or product. For example, a YouTuber might include a link to a sponsored product with a unique referral code. Drawing from the categorisation of monetisation models proposed by~\citet{goanta2019business}, we adapted and expanded these categories to reflect the diverse monetisation strategies on YouTube. We identified six main models:

\begin{enumerate}
    \item \textbf{Endorsements (Endors.)}: Creators promote a brand or service and receive compensation, such as payments or free products.
    \item \textbf{Affiliate Marketing (Aff.Mkt.)}: Creators include links with a unique code or referral link, receiving a commission for each sale or referral.
    \item \textbf{Direct Selling (Merch)}: YouTubers sell branded merchandise like t-shirts and accessories.
    \item \textbf{Subscription Services (Subs)}: Platforms like Patreon and OnlyFans, where followers pay to subscribe to access exclusive content.
    \item \textbf{Crowdfunding (Crowd.)}: Platforms for project funding (e.g., Indiegogo) or direct donations to creators (e.g., through PayPal links).
    \item \textbf{Others}: Monetisation sources that do not fit the previous categories.
\end{enumerate}

We primarily analysed URLs in video descriptions to systematically identify and categorise monetisation cues. We extracted all URLs and expanded them to their complete form. This expansion is necessary because many URLs are shortened for brevity; however, to accurately assess their content, especially for affiliate marketing links with unique identifiers for tracking sales and referrals, we must resolve these shortened URLs to their final destinations.

We identified 15,952 unique URLs, which collectively appeared 123,121 times in the data set, linked to 1,502 distinct canonical domains. We then ranked these domains by their occurrence frequency and manually classified the most common ones into our monetisation categories. Furthermore, we included a \textit{Self-promotion} category containing links to other social media platforms or content such as podcasts, which was the most frequent type of URL found in video descriptions.

Of these domains, we categorised 273, achieving a coverage of 92.89\% in terms of URL frequency. This metric indicates that our classification includes the majority of URLs, thus providing a comprehensive overview of the monetisation strategies employed. Moreover, we compiled a list of keywords associated with monetisation, expanding our analysis. The \textit{Other} category includes video descriptions that contain keywords related to monetisation, but do not necessarily contain a URL from the categorised list. The list of domains and keywords are available online\footnote{\url{https://github.com/thalesbertaglia/toxmon}}.

It is important to acknowledge potential overlaps between categories. For example, an Amazon URL with an affiliate code is classified under \textit{Affiliate Marketing}. However, if the same URL also contained the term \enquote{buy}, it was additionally tagged as \textit{Direct Selling} based on our keyword list. This approach aimed to maximise coverage rather than achieve high precision. \autoref{tab:monetization_metrics} presents the percentage of videos per channel and group that contain at least one monetisation cue within their descriptions for each monetisation category.

\begin{table}
\tiny
\centering
\caption{Percentage of videos with monetisation cues per channel.}
\label{tab:monetization_metrics}
\begin{tabular}{lcccccccc}
\toprule
\bfseries Channel & \bfseries Endors. & \bfseries Aff.Mkt. & \bfseries Merch & \bfseries Subs. & \bfseries Crowd. & \bfseries Other & \bfseries Any \\
\midrule
SSSniperWolf & 27.87 & 28.22 & 75.38 & 0.00 & 0.00 & 85.98 & 95.94 \\
James Charles & 60.30 & 61.60 & 65.49 & 0.00 & 0.00 & 39.52 & 66.23 \\
Logan Paul & 3.35 & 1.68 & 67.74 & 0.00 & 0.28 & 58.10 & 72.77 \\
Jake Paul & 0.35 & 57.67 & 62.72 & 0.00 & 0.70 & 70.30 & 74.56 \\
JennaMarbles & 0.00 & 0.80 & 100.00 & 0.00 & 0.00 & 3.60 & 100.00 \\
shane & 19.41 & 87.06 & 97.20 & 0.52 & 0.70 & 97.20 & 98.08 \\
David Dobrik & 0.19 & 7.28 & 49.07 & 0.00 & 1.49 & 39.74 & 51.12 \\
jeffreestar & 14.78 & 14.32 & 98.15 & 0.00 & 0.23 & 7.62 & 98.15 \\
Colleen Ballinger & 0.64 & 12.56 & 38.96 & 0.00 & 2.02 & 25.39 & 45.74 \\
The Gabbie Show & 0.00 & 26.53 & 51.02 & 6.12 & 0.00 & 44.90 & 59.18 \\
blndsundoll4mj & 21.11 & 38.63 & 63.79 & 3.41 & 9.87 & 47.45 & 68.38 \\
boogie2988 & 13.37 & 18.00 & 19.20 & 0.69 & 0.43 & 17.23 & 32.79 \\
Nikocado Avocado & 0.57 & 10.33 & 47.95 & 97.03 & 14.71 & 97.17 & 97.17 \\
The Completionist & 28.33 & 33.05 & 60.66 & 98.43 & 0.00 & 98.86 & 99.43 \\
iilluminaughtii & 57.19 & 55.13 & 32.23 & 11.06 & 0.00 & 66.03 & 91.63 \\
JessiSmiles & 3.16 & 8.42 & 20.53 & 0.00 & 0.00 & 22.11 & 27.89 \\
Yumi King & 5.88 & 46.88 & 51.10 & 23.44 & 26.01 & 34.38 & 67.74 \\
nickisnotgreen & 4.72 & 10.24 & 46.46 & 33.07 & 0.00 & 33.86 & 70.87 \\
Life Plus Cindy & 0.00 & 89.29 & 91.96 & 0.00 & 83.93 & 92.86 & 93.75 \\
lil lunchbox & 1.30 & 0.97 & 6.17 & 0.97 & 0.00 & 7.79 & 11.36 \\
\midrule
\textbf{Consistent} & 13.95 & 35.56 & 67.40 & 10.73 & 4.81 & 55.95 & 76.43 \\
\textbf{Spike} & 18.94 & 28.12 & 49.11 & 10.59 & 4.96 & 53.69 & 68.76 \\
\midrule
\bfseries All & 16.87 & 31.20 & 56.67 & 10.65 & 4.90 & 54.62 & 71.93 \\
\bottomrule
\end{tabular}
\end{table}

Most of the videos in our dataset (71.93\%) have monetisation cues, suggesting potential monetisation. \textit{Merch} is the most prevalent monetisation model, particularly for \textit{Consistent} channels, highlighting the effectiveness of merchandise sales as a monetisation strategy for most creators. On average, channels categorised as \textit{Consistent Controversy} are more likely to be monetised than those in the \textit{Spike} category. This difference may be explained by their larger followings, making them more attractive to brands and sponsorship opportunities. In particular, eight channels displayed monetisation cues in more than 90\% of their videos, indicating extensive monetisation despite potential controversies.

Building on our finding that a substantial portion of videos in our dataset contain at least one monetisation cue, we further quantify the extent of this monetisation. We measured both the total number of monetisation cues and their diversity across the dataset. The total number of cues indicates the intensity of monetisation efforts; essentially, it reflects the degree to which a creator emphasises monetisation strategies within their content. A higher total cue count suggests a more aggressive push toward monetising the video content. In contrast, the count of unique monetisation cues, determined by the number of unique URLs within each monetisation model, represents the breadth of monetisation strategies. This metric offers a view into the variety of brands, services, or platforms a creator collaborates with or uses for monetisation. \autoref{tab:num_monetisation_urls} presents the results. To exemplify the types of monetisation URLs from video descriptions in our dataset, \autoref{tab:top_mon_domains_by_category} shows the five most frequent domains of monetised URLs per category.

% \begin{landscape}
\begin{table}
\tiny
\centering
\caption{Number of monetisation cues across channels and categories. The numbers in brackets indicate the count of unique URLs identified within each monetisation category}
\label{tab:num_monetisation_urls}
\begin{tabular}{lccccccc}
\toprule
\bfseries Channel & \bfseries Endors. & \bfseries Aff.Mkt. & \bfseries Merch & \bfseries Subs. & \bfseries Crowd. & \bfseries Other \\
\midrule
SSSniperWolf & 1943 (26) & 821 (20) & 2906 (20) & 0 (0) & 0 (0) & 10 (10) \\
James Charles & 1380 (23) & 3 (3) & 403 (59) & 0 (0) & 0 (0) & 5 (4) \\
Logan Paul & 24 (1) & 2 (2) & 501 (15) & 0 (0) & 2 (2) & 4 (4) \\
Jake Paul & 4 (3) & 1827 (18) & 862 (22) & 0 (0) & 9 (4) & 44 (8) \\
JennaMarbles & 0 (0) & 0 (0) & 411 (6) & 0 (0) & 0 (0) & 0 (0) \\
shane & 111 (3) & 670 (20) & 733 (17) & 3 (1) & 7 (7) & 11 (9) \\
David Dobrik & 2 (1) & 5 (3) & 278 (10) & 0 (0) & 8 (5) & 2 (2) \\
jeffreestar & 107 (66) & 1 (1) & 794 (77) & 0 (0) & 1 (1) & 2 (2) \\
Colleen Ballinger & 8 (6) & 123 (10) & 772 (26) & 0 (0) & 23 (12) & 50 (8) \\
The Gabbie Show & 0 (0) & 10 (1) & 8 (2) & 3 (1) & 0 (0) & 2 (1) \\
blndsundoll4mj & 1222 (48) & 2227 (114) & 2788 (247) & 77 (6) & 217 (4) & 342 (36) \\
boogie2988 & 315 (25) & 120 (43) & 224 (28) & 16 (3) & 10 (8) & 19 (18) \\
Nikocado Avocado & 6 (3) & 65 (53) & 375 (18) & 1034 (3) & 136 (2) & 48 (20) \\
The Completionist & 218 (15) & 110 (18) & 695 (59) & 725 (4) & 0 (0) & 18 (15) \\
iilluminaughtii & 494 (56) & 196 (5) & 208 (13) & 70 (1) & 0 (0) & 73 (5) \\
JessiSmiles & 6 (6) & 6 (6) & 39 (9) & 0 (0) & 0 (0) & 2 (2) \\
Yumi King & 820 (745) & 859 (818) & 4168 (895) & 886 (40) & 299 (2) & 75 (70) \\
nickisnotgreen & 6 (5) & 1 (1) & 55 (3) & 42 (3) & 0 (0) & 5 (4) \\
Life Plus Cindy & 0 (0) & 556 (134) & 283 (68) & 0 (0) & 326 (2) & 3 (3) \\
lil lunchbox & 8 (7) & 2 (2) & 14 (6) & 3 (1) & 0 (0) & 5 (5) \\
\midrule
\bfseries Consistent & 2856 (148) & 4810 (215) & 7153 (470) & 1117 (10) & 380 (23) & 460 (86) \\
\bfseries Spike & 3818 (891) & 2794 (1057) & 9364 (1126) & 1742 (52) & 658 (24) & 260 (140) \\
\midrule
\bfseries All & 6674 (1039) & 7604 (1272) & 16517 (1596) & 2859 (62) & 1038 (47) & 720 (223) \\
\bottomrule
\end{tabular}
\end{table}
% \end{landscape}

% \begin{landscape}
\begin{table*}
\small
\centering
\caption{Five most frequent domains of monetised URLs per category across the entire dataset.}
\label{tab:top_mon_domains_by_category}
    \begin{tabular}{lcccccc}
    \toprule
    \bfseries  & \bfseries 1 & \bfseries 2 & \bfseries 3 & \bfseries 4 & \bfseries 5 \\
    \midrule
    \bfseries Subs. & patreon.com & trishafterdark.com & trishyland.com & onlyfans.com & n/a \\
    \bfseries Endors. & gtomegaracing.com & gfuel.com & shein.top & romwe.com & bellamihair.com \\
    \bfseries Aff. Mktg. & amazon.com & shein.com & ebates.com & gammagamers.com & go.com \\
    \bfseries Merch & amazon.com & spreadshirt.com & ogwolfpack.com & fanjoy.co & jeffreestarcosmetics.com \\
    \bfseries Crowdf. & cameo.com & buymeacoffee.com & paypal.me & paypal.com & gofundme.com \\
    \bfseries Other & sadboy2005.com & joinhoney.com & ggood.vip & cadoganhall.com & smarturl.it \\
    \bottomrule
    \end{tabular}
\end{table*}
% \end{landscape}

The number of monetisation cues reveals significantly different patterns between \textit{Consistent} and \textit{Spike} channels. Notably, \textit{Spike} channels have more monetisation URLs and a greater diversity of unique URLs, suggesting a greater effort to monetise their content than \textit{Consistent} channels. We hypothesise that these discrepancies may reflect distinct monetisation and marketing strategies. 

Certain channels, such as \textit{Yumi King}, include multiple affiliate links for various products, each representing, for instance, an item from a featured outfit. This approach is different from others that use a single promotional code. This method indicates a granular approach to monetisation, potentially increasing the likelihood of viewer engagement and purchase through direct product links, which might connect to the type of content or industry the channel promotes.

An interesting observation is Trisha Paytas's use of multiple versions of her merch website URL. Some of these links lead to no longer active pages, while others focus on different sets of products. This pattern suggests a self-moderation strategy (as discussed by~\citet{bertaglia2021clout}), where altering URLs serves to navigate around failed campaigns or controversial products. Switching URLs helps redirect attention away from less successful or problematic items, therefore, managing audience perception.

\begin{figure}[htbp]
    \centering
    \includegraphics[width=\linewidth]{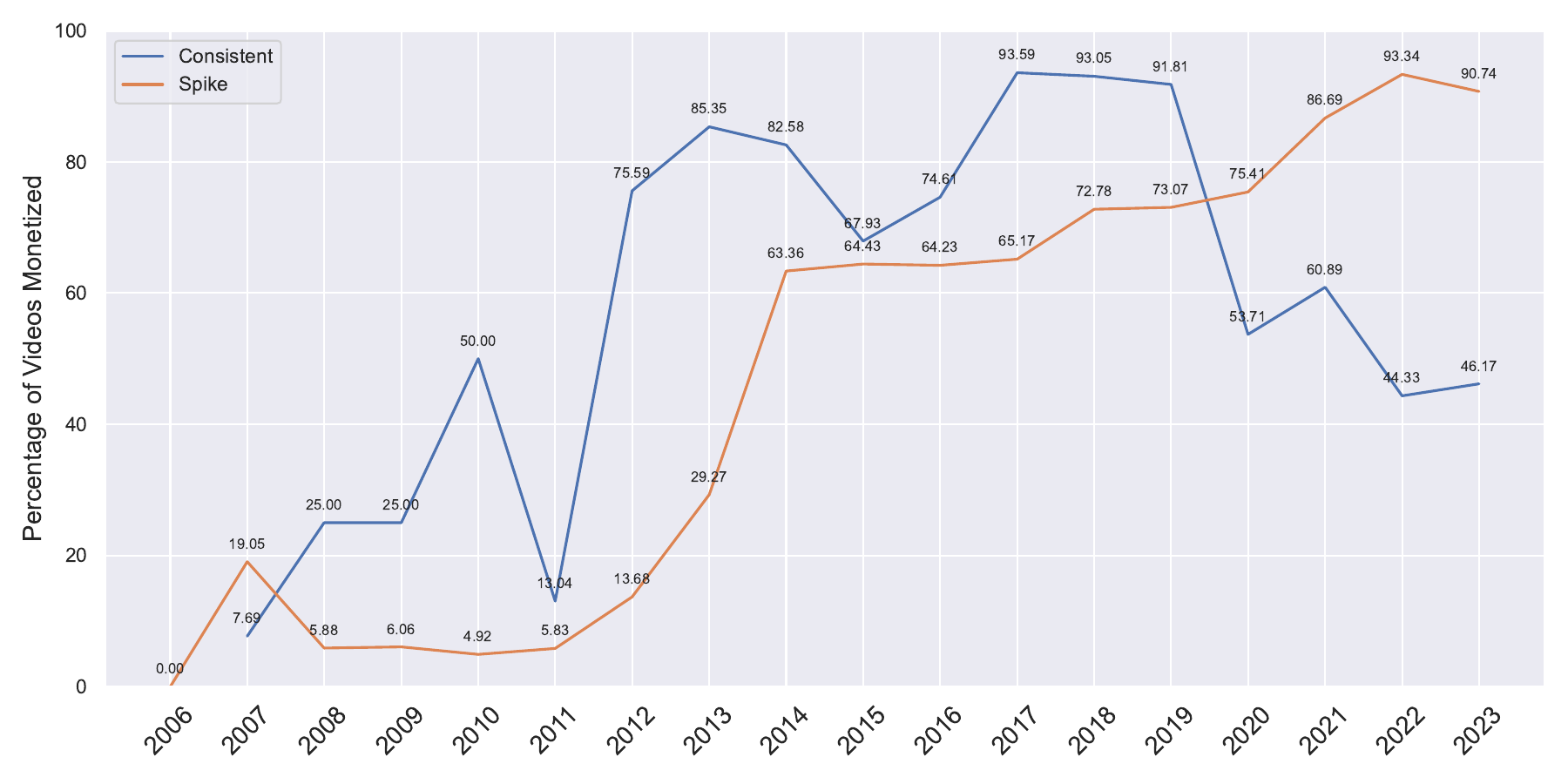}
    \caption{Percentage of videos with monetisation cues by channel category over time.}
    \label{fig:percentage_monetization_over_time}
\end{figure}

The difference in monetisation strategies might also stem from the channels' varying ages. \textit{Spike} channels, typically younger, may not have established long-term partnerships compared to \textit{Consistent} channels. Thus, \textit{Spike} channels might engage in a broader set of partnerships to diversify their monetisation efforts and explore different revenue streams. To further examine the temporal trends in monetisation, \autoref{fig:percentage_monetization_over_time} shows changes in video monetisation cues across different channel categories over time. 
% Furthermore, \autoref{fig:top_monetization_models_over_time} presents a detailed view of the percentage of videos featuring the top three monetisation models over time, enabling a focused analysis of how the most prevalent monetisation strategies have shifted and developed. 

% \begin{figure}[htbp]
%     \centering
%     \includegraphics[width=\linewidth]{figures/top_monetization_models_over_time}
%     \caption{Percentage of videos with the top three monetisation models over time.}
%     \label{fig:top_monetization_models_over_time}
% \end{figure}

\autoref{fig:percentage_monetization_over_time} shows a trend that aligns with the age of the channels. Initially, \textit{Consistent} channels, which include YouTubers who have been on the platform for a long time and were among the pioneers in monetising their content, had a higher number of videos with monetisation cues. However, as \textit{Spike} channels, generally newer to the platform, began to increase their activity, the trend started to even out. Monetisation on \textit{Spike} channels consistently increased, surpassing \textit{Consistent} channels from 2019.

A significant decline in the monetisation cues of \textit{Consistent} channels around the same time further accentuated this trend. This drop could be linked to several high-profile creators within this group becoming less active on YouTube, possibly due to moving to other platforms or the impact of controversies leading to them being \enquote{cancelled}~\cite{lewis2022platform} and consequently reducing their activity. A notable example is Shane Dawson, who faced widespread backlash and controversy over past content and actions, leading to a hiatus from the platform around 2020~\cite{geusens2023cancelled}. This case exemplifies how controversies can affect a creator's activity and presence on YouTube, directly impacting their ability to monetise content. The timing of these events correlates with the observed decrease in monetisation cues, illustrating the potential vulnerability of established creators to shifts in public perception and platform dynamics.

\section{Measuring Toxicity}
\label{sec:measuring-tox}
Following our exploration of monetisation cues, we focus on analysing the toxicity of comments to understand its relationship with controversial content. To quantitatively measure toxicity, we train a regression model to predict toxicity scores based on the content of the comments. We use the ALYT dataset, which comprises YouTube comments from controversial videos labelled for abusive language detection on a scale of 1 to 7~\cite{bertaglia2021abusive}. This data set matches the context of our study and provides a nuanced understanding of toxicity levels through its thorough annotation methodology. 

Considering the massive volume of our comment dataset, including over 100 million comments, selecting a simple and efficient machine model was essential. We used a Ridge regression model with tf-idf features to balance computational efficiency and predictive accuracy. Ridge regression is a type of linear regression with a regularisation term that penalises overly complex models that overfit training data and underperform on unseen data~\cite{mcdonald2009ridge}; we use scikit-learn's implementation with default hyperparameters~\cite{scikit-learn}. 

Although Perspective API~\cite{lees2022new} is a popular tool widely used in many studies to detect and measure toxicity, we opted not to use it in our analysis for several reasons. Firstly, the API has been shown to exhibit biases and present various pitfalls, which can lead to inaccurate or skewed toxicity scores~\cite{rosenblatt2022critical}. These issues are particularly concerning when studying diverse and large-scale datasets, where biases in toxicity detection could significantly impact the results. Secondly, the API's rate limits make it impractical to process our dataset, which includes over 100 million comments. Given the size and scope of our data, the rate limitations of the Perspective API would have greatly hindered the efficiency of our analysis. Consequently, we chose to train our own Ridge regression model with tf-idf features, balancing computational efficiency with predictive accuracy, and ensuring that our approach was better suited to the scale and complexity of our dataset.

We prepared the ALYT dataset for model training by removing comments labelled with a 4, which indicated annotator uncertainty. We removed 519 comments, leaving 19,396 comments in the remaining training dataset. We normalised the labels to a continuous scale using a MinMaxScaler to ensure that our regression model could interpret the data effectively. Additionally, we preprocessed the comments by removing multiple consecutive spaces and converting them to lowercase. Following the preprocessing step, we extracted tf-idf features with ngrams ranging from 1 to 3.

To validate the model's ability to predict toxicity scores accurately, we conducted a 10-fold cross-validation on the ALYT dataset, evaluating its performance with several metrics: Mean Squared Error (MSE), Root Mean Squared Error (RMSE), Mean Absolute Error (MAE), and the coefficient of determination ($R^2$). These metrics respectively measure the average square difference between the estimated values and actual value, the square root of MSE providing a scale similar to the original data, the average absolute difference, and the proportion of the variance in the dependent variable (toxicity scores) that is predictable from the independent variables (features extracted from comments)~\cite{botchkarev2018performance,chicco2021coefficient}. \autoref{tab:tox_cv} presents the cross-validation results.

\begin{table}
\centering
\caption{Cross-validation results for our toxicity prediction model, including mean and standard deviation calculated across ten folds, and maximum and minimum values for MSE, RMSE, MAE, and $R^2$.}
\label{tab:tox_cv}
    \begin{tabular}{@{}lllll@{}}
    \toprule
                   & \multicolumn{1}{c}{\textbf{Mean}} & \multicolumn{1}{c}{\textbf{Std}} & \multicolumn{1}{c}{\textbf{Max}} & \multicolumn{1}{c}{\textbf{Min}} \\ \midrule
    \textbf{MSE}   & 0.031                             & 0.001                            & 0.032                            & 0.029                            \\
    \textbf{RMSE}  & 0.175                             & 0.003                            & 0.177                            & 0.170                            \\
    \textbf{MAE}   & 0.129                             & 0.002                            & 0.132                            & 0.127                            \\
    \textbf{$R^2$} & 0.381                             & 0.015                            & 0.408                            & 0.356                            \\ \bottomrule
    \end{tabular}
\end{table}

The results indicate a reasonable performance of our model, especially reflected in the RMSE value of 0.175. Considering the scale of toxicity scores ranges from 0 to 1, an RMSE of 0.175 shows that the model's predictions are, on average, within 0.175 units of the actual toxicity scores. Furthermore, the value $R^2$ of 0.381 indicates that the model can explain approximately 38.1\% of the variance in toxicity scores. Although this result suggests that the model captures some predictive relationships, it also highlights that there is room for improvement, as a significant portion of the variance remains unexplained. The relatively low RMSE and MSE values indicate that the model predictions are not far from the observed values, demonstrating some level of predictive accuracy. However, it is likely that the model does not fully capture all variables or relationships in the data. Future work could explore alternative models or more complex approaches to potentially improve model performance. Given these considerations, we trained the model on the entire ALYT dataset and used it to predict the toxicity scores of the comments in our dataset.

The predicted toxicity scores indicate that most comments are non-toxic, reflecting relatively low scores, which aligns with the effects of YouTube content moderation practices. However, relying only on the average toxicity score to aggregate video toxicity tended to produce uniformly low values, offering limited insights into the nuances of actual comment toxicity. To address this challenge, we implemented a post-processing technique to refine the aggregation of raw toxicity scores provided by the model. This method aims to capture a more nuanced view of toxicity by considering the distribution of scores rather than just their average. Specifically, we calculate an adjusted aggregate toxicity score for each video, incorporating measures of dispersion and central tendency from the comments' toxicity scores:
$$
\text{Adjusted Score} = \text{Median Toxicity} + (Q_{75\%} \text{Toxicity} - Q_{25\%} \text{Toxicity})
$$

Our Adjusted Toxicity Score combines the median toxicity (representing the central trend) with the interquartile range (IQR, calculated as $Q_{75\%}$ Toxicity - $Q_{25\%}$ Toxicity), which measures the dispersion of toxicity scores around the median. This approach acknowledges that the presence of highly toxic comments (reflected by a wider IQR) can significantly impact the perceived toxicity of a video's comment section, even if the median toxicity score is low. By adjusting for the IQR, we aim to provide a more comprehensive measure of aggregated video toxicity that accounts for both the typical (median) comment toxicity and the variability (dispersion) of toxicity levels within comments. We follow the same approach when aggregating the video-level toxicity scores to determine the channel-level toxicity. All aggregate results we present use the adjusted toxicity score; we also multiply the scores by one hundred to improve readability. \autoref{tab:channel_tox_stats} shows the aggregated toxicity metrics across channels and categories.

\begin{table}
\small
\centering
\caption{Aggregated toxicity metrics across channels and categories: Toxicity scores are represented by mean and standard deviation values. The range column displays the lowest and highest toxicity scores within each group. The High Toxicity column indicates the percentage of videos whose toxicity score surpasses the average toxicity score for the entire dataset.}
\label{tab:channel_tox_stats}
    \begin{tabular}{lcccc}
    \toprule
    \bfseries Channel & \bfseries Toxicity & \bfseries Range & \bfseries High Toxicity (\%) \\
    \midrule
    SSSniperWolf & 31.43$\pm$3.18 & 19.15-46.16 & 35.88 \\
    James Charles & 27.86$\pm$2.42 & 17.61-36.06 & 13.08 \\
    Logan Paul & 32.90$\pm$3.79 & 19.85-54.89 & 60.13 \\
    Jake Paul & 34.39$\pm$4.33 & 20.00-54.46 & 66.63 \\
    JennaMarbles & 32.33$\pm$2.84 & 22.54-40.13 & 58.33 \\
    shane & 32.40$\pm$3.03 & 23.14-54.63 & 67.46 \\
    David Dobrik & 31.50$\pm$2.01 & 24.47-38.78 & 74.34 \\
    jeffreestar & 29.54$\pm$2.25 & 21.59-37.66 & 27.78 \\
    Colleen Ballinger & 28.07$\pm$2.56 & 18.01-37.96 & 11.14 \\
    The Gabbie Show & 32.01$\pm$3.13 & 22.63-37.92 & 35.90 \\
    blndsundoll4mj & 33.41$\pm$3.99 & 13.83-51.50 & 48.28 \\
    boogie2988 & 36.38$\pm$4.47 & 17.55-62.42 & 66.68 \\
    Nikocado Avocado & 35.17$\pm$4.60 & 19.96-50.62 & 57.61 \\
    The Completionist & 29.31$\pm$1.80 & 20.43-32.57 & 25.18 \\
    iilluminaughtii & 34.09$\pm$2.99 & 21.46-39.43 & 70.93 \\
    JessiSmiles & 32.80$\pm$4.94 & 18.43-55.79 & 42.25 \\
    Yumi King & 25.99$\pm$2.77 & 14.35-34.74 & 4.51 \\
    nickisnotgreen & 32.51$\pm$3.44 & 18.53-37.32 & 62.20 \\
    Life Plus Cindy & 26.48$\pm$3.46 & 15.00-35.86 & 9.46 \\
    lil lunchbox & 41.55$\pm$7.79 & 12.33-67.25 & 82.94 \\
    \midrule
    \bfseries Consistent & 33.15$\pm$3.89 & 13.83-54.89 & 52.71 \\
    \bfseries Spike & 32.74$\pm$4.45 & 12.33-67.25 & 40.84 \\
    \midrule
    \bfseries All & 32.99$\pm$4.25 & 12.33-67.25 & 45.67 \\
    \bottomrule
    \end{tabular}
\end{table}

The results show a uniform average toxicity level between \textit{Consistent} and \textit{Spike} channels, suggesting that toxic comments are common across the dataset. However, \textit{Spike} channels have higher peaks of toxicity, with \textit{Lil Lunchbox} standing out as an outlier; this channel exhibits the highest average (41.55) and maximum (67.25) toxicity scores and also a significant percentage (82.94\%) of videos with above-average toxicity. The presence of outlier channels with extreme toxicity scores indicates that specific content or topics can cause intense reactions, exceeding the usual toxicity observed across the dataset. Moreover, the association between higher standard deviations and elevated average toxicity levels further suggests that specific videos with spikes in toxicity contribute to the overall increase in a channel's average toxicity, pointing to particular videos as potential sources of controversy. 

Conversely, channel \textit{The Completionist} shows lower than expected toxicity levels, possibly due to the recency of the channel's controversy, which unfolded in the same month as our data collection. This limited time frame could explain the absence of a significant increase in toxic comments, suggesting that it might take longer for audience reactions to occur. This scenario highlights a critical insight: responses to controversies on YouTube can be delayed, with spikes in toxicity often occurring in response to previous videos rather than immediately following a contentious event. 

\section{Toxicity, Engagement and Monetisation}
Finally, we analyse the relationship between toxicity, engagement, and monetisation to understand how toxicity affects engagement and monetisation and, by extension, how controversy influences these metrics. Given these complex and dynamic relationships, influenced by multiple factors not captured within our dataset, our analysis is inherently constrained by the variables we can observe. Nonetheless, we employ regression analysis to find direct and noticeable associations between these variables, fully acknowledging the limitations of this approach in capturing the vast scope of potential interactions.

We perform a regression analysis using ordinary least squares to determine whether increases or decreases in toxicity correlate with changes in engagement metrics (view counts, comment counts, and like counts) and the prevalence of monetisation cues, measured as the total of monetisation cues within video descriptions. We normalise engagement metrics to a log-scale to account for their skewed distribution. \autoref{tab:tox_reg} presents the regression results, including the coefficient (indicating the nature and strength of the relationship), the standard error (measuring the precision of the coefficient) and the 95\% confidence interval (providing a range within which the coefficient is likely to fall). Values marked with *** denote statistically significant relationships, with $p < 0.001$, highlighting associations with high confidence.

\begin{table}[htbp]
\caption{Ordinary least squares regression results using toxicity as the predictor variable. *** indicates $p < 0.001$.}
\label{tab:tox_reg}
% \small
\centering
\begin{tabular}{@{}lccc@{}}
\toprule
\multicolumn{1}{c}{\textbf{Feature}} & \textbf{Coefficient} & \textbf{Std Err} & \textbf{CI (95\%)} \\ \midrule
\textbf{Views (Log)}                   & -0.3613            & 0.3956                   & {[}-1.1367 0.4141{]}                   \\
\textbf{Comments (Log)}                    & 3.1143***           & 0.3248                   & {[}2.4776 3.7510{]}                 \\
\textbf{Likes (Log)}               & -2.6013***           & 0.3838                   & {[}-3.3537 -1.8490{]}                 \\
\textbf{Monetisation Cues}                 & -8.7149***            & 0.3373                   & {[}-9.3762 -8.0537{]}                   \\ \bottomrule
\end{tabular}
\end{table}

The analysis shows that the connection between toxicity and viewer interaction is not statistically significant; thus, these results are inconclusive. In contrast, a significant positive correlation with comments indicates a connection between increased toxicity and viewer interaction. This connection suggests that while controversial content may provoke more discussions, this engagement could be from negative interactions. The negative relationship between toxicity and likes further illustrates this, indicating that more toxic content tends to receive fewer likes, reflecting a decline in overall viewer approval. Critically, the pronounced negative effect of toxicity on monetisation underscores a critical insight: videos with higher toxicity levels generally have fewer monetisation cues. Despite potentially higher engagement rates, controversial content does not necessarily correlate with the financial benefit of content creators. This scenario suggests that controversy can lead to more engagement, but it might not be a viable strategy for increasing monetisation.

It is important to note that these results can be significantly influenced by channels with very high average toxicity, such as \textit{Lil Lunchbox}, which generally have fewer monetisation cues. Furthermore, the potential delayed effect of negative reactions, as explored in \autoref{sec:measuring-tox}, could play a role in these dynamics, indicating that the impact of controversy on monetisation and engagement could evolve over time.

\section{Summary}
In conclusion, our analysis of controversial YouTube content discusses the complex interplay between toxicity, engagement, and monetisation within the creator economy. Our findings indicate that while toxicity is linked to higher levels of viewer interaction -- most notably, an increase in comment volume -- such engagement often has negative implications, leading to fewer likes and monetisation cues. However, it is crucial to highlight that outlier channels with exceptionally high levels of toxicity could skew these results. Moreover, the potentially delayed responses to toxic content indicate that the impact of controversy on engagement and monetisation may evolve gradually.

Our study also uncovers evidence of self-moderation practices among content creators, such as using various URLs to sell merchandise to potentially circumvent the repercussions of failed campaigns or controversial products. Moreover, we observed a significant variation in the range of monetisation URLs among certain controversial YouTubers, possibly indicating reliance on numerous, potentially lower-quality sponsorships or challenges in maintaining long-term partnerships. However, our current analysis, primarily focused on the number of monetisation cues, does not fully address these nuances. Future research should delve deeper into the qualitative aspects of monetisation, aiming to understand the specifics of what controversial YouTubers promote, particularly focusing on potentially harmful or sensitive categories like financial services, health products, and gambling websites. Additionally, understanding the delayed effects of toxicity would be a relevant research direction.

Our findings have practical implications for multiple stakeholders in the YouTube ecosystem, including content creators, platform regulators, and advertisers. For content creators, understanding the relationship between controversy, toxicity, and monetisation can guide more informed decisions about content strategies, balancing the pursuit of engagement with potential negative impacts on audience sentiment and monetisation opportunities. Platform regulators can use these insights to develop more nuanced content moderation policies, potentially targeting high-toxicity content that might drive engagement but harm the overall community experience. Finally, advertisers could leverage this knowledge to refine their targeting strategies, avoiding association with highly toxic content that might undermine brand reputation while aligning with creators whose content promotes healthier community interactions. Overall, our study offers a foundation for future research and practical applications to improve the dynamics of social media platforms.

\begin{acks}
    This research has been supported by funding from the ERC Starting Grant HUMANads (ERC-2021-StG No 101041824).
\end{acks}
%%
%% The next two lines define the bibliography style to be used, and
%% the bibliography file.
\bibliographystyle{ACM-Reference-Format}
\bibliography{main}

%%
%% If your work has an appendix, this is the place to put it.
% \appendix

% \section{Research Methods}

% \subsection{Part One}

% Lorem ipsum dolor sit amet, consectetur adipiscing elit. Morbi
% malesuada, quam in pulvinar varius, metus nunc fermentum urna, id
% sollicitudin purus odio sit amet enim. Aliquam ullamcorper eu ipsum
% vel mollis. Curabitur quis dictum nisl. Phasellus vel semper risus, et
% lacinia dolor. Integer ultricies commodo sem nec semper.

% \subsection{Part Two}

% Etiam commodo feugiat nisl pulvinar pellentesque. Etiam auctor sodales
% ligula, non varius nibh pulvinar semper. Suspendisse nec lectus non
% ipsum convallis congue hendrerit vitae sapien. Donec at laoreet
% eros. Vivamus non purus placerat, scelerisque diam eu, cursus
% ante. Etiam aliquam tortor auctor efficitur mattis.

% \section{Online Resources}

% Nam id fermentum dui. Suspendisse sagittis tortor a nulla mollis, in
% pulvinar ex pretium. Sed interdum orci quis metus euismod, et sagittis
% enim maximus. Vestibulum gravida massa ut felis suscipit
% congue. Quisque mattis elit a risus ultrices commodo venenatis eget
% dui. Etiam sagittis eleifend elementum.

% Nam interdum magna at lectus dignissim, ac dignissim lorem
% rhoncus. Maecenas eu arcu ac neque placerat aliquam. Nunc pulvinar
% massa et mattis lacinia.

\end{document}